\documentclass[10pt,leqno]{article}
\usepackage{graphicx}
\usepackage{indentfirst,csquotes}
\usepackage{natbib}

\RequirePackage{geometry}
\usepackage{amssymb,amsthm,amsmath}
\usepackage{xcolor,paralist,hyperref,titlesec,fancyhdr,etoolbox}
\usepackage{mdframed} 

\usepackage{fancyhdr}  
\fancypagestyle{plain}{\fancyhf{}\fancyfoot[R]{\thepage}}

\renewenvironment{abstract}{%
    \smallskip
    \begin{center}
    \begin{minipage}{0.9\textwidth} 
    \small
}{%
    \end{minipage}
    \end{center}
    \bigskip
}

\usepackage{lipsum}
\hypersetup{
    pdfborder={0 0 0}  
}

\begin{document}
\title{Hopper flows of dense suspensions: a 2D microfluidic model system}
\author{Lars Kool\textsuperscript{1,2,*}, Jules Tampier\textsuperscript{1,*}, Philippe Bourrianne\textsuperscript{1,\dag}, Anke Lindner\textsuperscript{1,3,\ddag}}  
\renewcommand{\MakeUppercase}[1]{#1} 
\date{}

\maketitle

\vspace{-.8cm} 
\begin{center}
    \small
    \textsuperscript{1}Laboratoire de Physique et Mécanique des Milieux Hétérogènes (PMMH), UMR 7636 CNRS - ESPCI Paris - Sorbonne Université – Université Paris Cité, France \\
    \textsuperscript{2}IPGG, 6 rue Jean-Calvin, 75005 Paris, France \\
    \textsuperscript{3}Institut Universitaire de France (IUF) \\
    \textsuperscript{*}These authors contributed equally to this work.\\
    \textsuperscript{\dag}Corresponding author. E-mail:
\thanks{philippe.bourrianne@espci.fr}\\
    \textsuperscript{\ddag}Corresponding author. E-mail: \thanks{anke.lindner@espci.fr}
\end{center}

\vspace{-.5cm} 
\begin{abstract}
Flows of particles through bottlenecks are ubiquitous in nature and industry, involving both dry granular materials and suspensions. However, practical limitations of conventional experimental setups hinder the full understanding of these flows in confined geometries. Here, we present a microfluidic setup to investigate experimentally the flow of dense suspensions in a two-dimensional hopper channel. Particles with controlled properties are in-situ fabricated with a photolithographic projection method and compacted at the channel constriction using a Quake valve. The setup is characterized by examining the flow of a dense suspension of hard, monodisperse disks through constrictions of varying widths. We demonstrate that the microfluidic hopper discharges particles at constant rate, resulting from the channel resistance being dominated by the presence of densely packed particles within the tapered section of the hopper. Under imposed flow rate the discharge remains independent of particle and orifice sizes, whereas it exhibits a Beverloo-like scaling under pressure-imposed conditions. Additionally, we show that the statistics of clog formation in our microfluidic hopper follow the same stochastic laws as reported in other systems. Finally, we show how the versatility of our microfluidic model system can be used to investigate the outflow and clogging of suspensions of more complex particles.
\end{abstract} 

\vspace{-.5cm} 
\begin{mdframed}
\textbf{\mathversion{bold}Impact Statement}
Flows of dry and wet granular suspensions through constrictions have long been studied from the regular discharge of sand in antique hourglasses to the granular flows within silos or even the clogging of particles under confinement. We here describe a novel and versatile 2D microfluidic hopper to investigate the flow and clogging dynamics of dense suspensions. The setup combines an in-situ microfluidic fabrication technique with a pneumatic valve at the orifice to print and concentrate particles of controllable shape. We here demonstrate that the microfluidic flow and clogging of dense suspensions of hard monodisperse disk-shaped particles are reminiscent of well-established behaviors in conventional granular hoppers. By conducting experiments under different driving forces, we report a constant particle discharge rate that depends on the flow driving mechanism and follows well-established stochastic clogging laws. Such results might contribute to bridge the gap between dry granular media and dense wet suspensions. Finally, the versatility and tunability of the microfluidic hopper pave the way for systematic experimental studies on complex granular suspensions involving particles of varying shapes and deformability, of obvious interest in practical applications.
\end{mdframed}

\newpage

\section{Introduction}

Particulate flows through bottlenecks are observed across a wide range of situations, both in nature and industry. 
These flows can involve dry granular materials, such as in grain silos \citep{jenike_quantitative_1967} and hourglasses \citep{pongo_flow_2021}, or particles suspended in a viscous fluid, as found in blood vessels \citep{patnaik_vascular_1994} and ink-jet printing processes \citep{croom_mechanics_2021}.
Research on particulate flow through a narrow orifice dates back to the mid- 1800's, with early studies focusing on the gravity-driven discharge of dry grains from silos \citep{hagen1852druck, janssen1895versuche}. Unlike liquids flowing from a container, grains exit a silo at a constant rate, independent of the filling height. While this constant discharge rate has long been attributed to pressure saturation at the bottom of the silo \citep{tighe_pressure_2007, sperl2006experiments}, more recent studies suggest that it is instead due to the constant exit velocity of particles in the outlet region \citep{aguirre_pressure_2010}.

The dependence of the constant particle discharge rate on the orifice and particle sizes in granular gravity-driven flows is commonly described by the Beverloo law \citep{beverloo_flow_1961}. While some of the physical concepts underlying this law have been questioned in recent studies \citep{mankoc_flow_2007, janda_flow_2012, rubio-largo_disentangling_2015, zhou_gas-assisted_2019}, its scaling remains universally observed across a wide range of situations \citep{nedderman_flow_1982}. It can even be applied to systems with fundamentally different driving mechanisms, including particles floating at the surface of a flowing fluid \citep{guariguata_jamming_2012} or transported on a conveyor belt moving at constant velocity \citep{aguirre_pressure_2010}.
When the orifice size becomes sufficiently small relative to the particle diameter, flow can be impeded by the formation of self-supporting arches at the constriction, leading to temporary or permanent interruptions of the discharge process. This phenomenon, observed in both dry systems \citep{zuriguel_invited_2014} and suspension flows \citep{marin_clogging_2024}, can significantly impact the performance of various industrial processes. 

Both the discharge and clogging of particles through a narrow orifice are controlled by a large number of parameters, including the particle size \citep{beverloo_flow_1961, janda_jamming_2008}, shape \citep{ashour_outflow_2017}, polydispersity \citep{govender_study_2018}, deformability \citep{tao_soft_2021} and solid fraction \citep{vani_influence_2022}, as well as the geometry of the channel \citep{genovese_crystallization_2011, vani_role_2024}. Among these parameters, the presence of a liquid phase significantly influences the discharge of particles through a constriction, due to fluid-particle interactions and lubrication effects \citep{guazzelli_physical_2011}. For instance, studies on the gravity-driven discharge of submerged granular hoppers revealed non-constant particle outflow rates, in contrast to dry silos \citep{wilson_granular_2014, koivisto_friction_2017}. The coupling between particles and the interstitial fluid can also influence the clogging of microfluidic systems \citep{dressaire_clogging_2016}, where the stability of clogging arches depends on the flow driving mechanism \citep{souzy_role_2022}.
Despite their importance, experimental studies on dense suspension flows through constrictions remain relatively scarce. One reason for this is the practical difficulty of working with dense suspensions, as jamming can occur at any point within the experimental system \citep{wyss_mechanism_2006}. As a result, experiments often start with more dilute suspensions, relying on clog formation to concentrate particles near the orifice \citep{dressaire_clogging_2016}, limiting the study of smaller particles which rarely clog.

In this paper, we present an experimental setup designed to study the flow of suspensions through a bottleneck in a two-dimensional configuration. Experiments are performed in a microfluidic hopper channel, where in-situ fabricated particles are transported through a constriction by a viscous fluid. While our experimental setup allows for a precise control over particle shape, deformability and volume fraction, here, only the flow of densely packed, hard, disk-shaped particles is studied. We demonstrate that, despite fundamental differences with dry granular silos, our experimental system discharges at a constant rate under both imposed flow rate and pressure conditions. Furthermore, we show that the discharge rate of particles in our microfluidic hopper follows a Beverloo-like scaling under pressure-imposed conditions, and that the stochastic laws governing particle clogging are recovered. This makes our setup an effective model system for studying particulate flow through bottlenecks across a wide range of configurations. 
The paper is organized as follows: We first present the experimental setup and particle fabrication method, and characterize the driving force in our system. We, then, examine the discharge of a dense packing of hard disk-shaped particles, and apply our system to the study of the hopper discharge and the clogging of particles. Finally, we give perspectives for future work involving more complex particle suspensions.

\section {Materials and Methods}
\subsection{Experimental setup}

\begin{figure}[tb]
    \begin{center}
        \includegraphics[width=.79\linewidth]{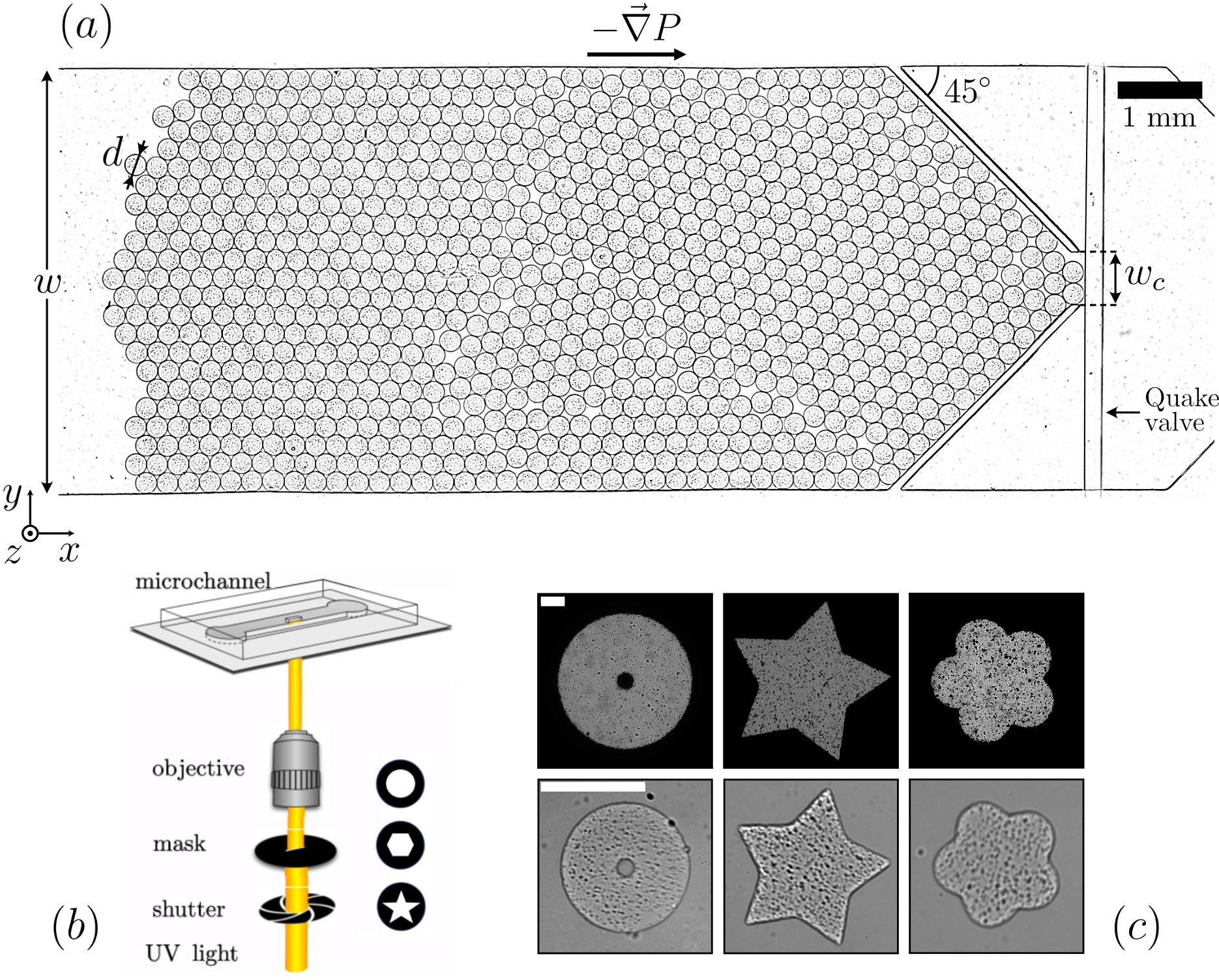}
        \caption{(a) Top view of the microfluidic hopper channel used in the experiments. The channel consists of a $24$ mm long straight section of fixed width $w=5$ mm, followed by a tapered section with an angle of 45\textdegree and a constriction width $w_c$. Particles of diameter $d$ are initially packed at the channel constriction using a Quake valve, before being discharged through the constriction. (b) Principle of particle fabrication. A UV curable polymer solution is filled into the microfluidic channel and illuminated through a mask, resulting in hydrogel-like particles fabricated inside the channel. Image adapted from \cite{cappello_transport_2019}. (c) Examples of masks and the corresponding crosslinked particles. Both scale bars are $200$ \textmu m.} 
        \label{fig:method}
    \end{center}
\end{figure}

We experimentally study the discharge of a dense suspension of hard disk-shaped particles through a two-dimensional microfluidic hopper. Both particle fabrication and discharge experiments are conducted in a $24$ mm long PDMS channel with fixed width $w=5$ mm and thickness $H=92$ \textmu m, as depicted in Figure~\ref{fig:method}(a). The channel features a hopper geometry, with a constriction of varying width $w_c=600$ or $900$ \textmu m, formed by a tapered section with an angle of $45$\textdegree. A pneumatically controlled Quake valve \citep{unger_monolithic_2000} is placed immediately after the constriction, retaining particles while allowing the interstitial fluid to pass when closed. Due to its small thickness relative to its width ($H\ll w$), the microfluidic channel exhibits a Hele-Shaw geometry. The velocity profile is thus parabolic in the $z$-direction while being independent of the $x$ and $y$ positions, except very close to the side walls and in the tapered section.

The microfluidic device consists of three layers (bottom to top): a glass microscopy slide spin coated with PDMS, a thin PDMS slab containing the hopper channel, and a thick PDMS slab containing the Quake-valve. The two first layers are fabricated by spin coating a $200 - 300$ \textmu m layer of PDMS (Sylgard 184, 19:1 base:crosslinker) onto a glass slide and a hopper channel mold ($92$ \textmu m SU-8 layer on a silicon wafer), while the third layer is obtained by casting a Quake-valve mold ($\sim 50$ \textmu m SU-8 layer on a silicon wafer) in $\sim$5 mm of PDMS (Sylgard 184, 9:1). All three layers are then cured separately at 70 \textdegree C for 40 minutes, after which the Quake-valve is bonded to the hopper channel, and the assembly is bonded to the spin-coated glass slide. Bonding is achieved by pressing the layers together and placing the assembly in an oven at 70 \textdegree C for 30 minutes \citep{kool_microfluidic_2023}.

The microfluidic hopper channel is observed using an inverted microscope (Zeiss Axio Observer A1), installed on an optical breadboard (Newport SG breadboard), passively leveled and isolated from vibrations by pneumatic feet (Newport CM-225). The microscope has a UV-light (Osram HBO 103W/2), filtered using a band-pass filter at $\lambda = 365\pm 10$ nm (Chroma D365/10), and reflected into the channel using a reflector cube (Zeiss 424933). The filtered UV-light is modulated precisely by an electronic shutter (Uniblitz V25, $100$ Hz), coupled to an external signal generator (Agilent A33220A). The channel position is modified using a motorized stage (Märzhäuser Wetzlar Tango PCI-E), which can be operated manually, or digitally via a LabView program. Fluid flow in the microfluidic channel is controlled either by imposing a pressure gradient $\Delta P$ with a pressure controller (Fluigent Flow EZ), or by setting a flow rate $Q$ with a syringe pump (Nemesys, Cetoni). For pressure-driven experiments, the tubing length connecting the fluid reservoir to the channel is systematically measured and kept constant to maintain the hydraulic resistance consistent across experiments. Movies of the experiments are recorded at $5$ Hz using a digital CMOS camera (Hamamatsu Orca-Fusion C14440).

\subsection{Particle fabrication}
\label{sec:fabrication}


The particles used in our experiments are fabricated using an in-situ photo-lithographic projection method \citep{dendukuri_stop-flow_2007, cappello_transport_2019}, as described in Figure~\ref{fig:method}(b). The microfluidic channel is first filled with a UV-polymerizable solution, whose composition depends on the desired deformability of the particles \citep{duprat_microfluidic_2014}. In this paper, we limit ourselves to a mixture of $90$\% Poly(ethylene glycol) diacrylate (PEGDA, $M_N$ = 575) and $10$\% photo-initiator (Darocur 1173), which results in crosslinked particles with a Young's modulus of $\sim$12 MPa \citep{cappello_transport_2019, duprat_microfluidic_2014}, rendering deformations negligible under our experimental conditions. A small amount of surfactant (0.1 \% v/v of Tween-80) is added to the polymer solution to mitigate adhesive interactions between particles. The solution is Newtonian of viscosity $\eta=47$ mPa.s and density $\rho=1120$ kg.m$^{-3}$.

Once the channel is filled with the UV-sensitive solution, fluid flow is stopped and the channel is locally exposed to UV light, resulting in the formation of a solid hydrogel-like particle. The crosslinked particle has the same density as the uncrosslinked solution and is considered frictionless, since hydrogels have a notoriously low friction coefficient (typically $10^{-3}$ \citep{gong2006friction}). Its shape and size are controlled by placing a mask in the field-stop position of the microscope. Examples of masks and the corresponding crosslinked particles are given in Figure~\ref{fig:method}(c). In this study, we only focus on disk-shaped particles with diameters $d$ ranging from $114$ to $378$ \textmu m. Due to the presence of oxygen near the channel walls, the crosslinking reaction is inhibited over a layer of thickness $\delta$ near the top and bottom walls of the channel \citep{dendukuri2008modeling}. This leads to particles with a thickness $h=H-2\delta$, where $\delta$ was estimated to be $21$ \textmu m in our experiments, yielding $h=50$ \textmu m. This thickness is sufficiently large to prevent overlapping and maintain a single layer of particles within the channel.

The microfluidic channel is filled with particles by repeating the aforementioned UV-crosslinking and moving the focal spot of the objective between successive UV pulses, enabling precise control over the microscopic particle configuration and initial packing fraction (see Supplementary Movie 1). Since this study focuses on dense packings, particles are fabricated in a hexagonal configuration to maximize their number within the channel. Note that particles must be separated by a few microns during crosslinking to avoid the fusion of closed structures, limiting the maximum packing that can be obtained. Once the fabrication is complete, particles are thus densely packed at the constriction of the hopper by closing the Quake valve and imposing a fluid flow in the channel (see Supplementary Movie 2).

\subsection{Hopper discharge experiment}

After the particles are densely packed at the constriction, the Quake valve is opened and the suspension is discharged through the channel orifice (see Supplementary Movie 3). If a clog forms, a brief reverse flow is applied to break the obstructing arch and restore particle flow (see Supplementary Movie 4). Each experiment concludes when the number of remaining particles is insufficient to fill the tapered section of the channel.

The recorded movies are then analyzed to detect particles and reconstruct their trajectories. For each frame, particles are detected using a fully parallelized in-house convolutional algorithm, with a precision of $\sim0.3$ \textmu m. The particle positions are then linked over time using the open-source tracking algorithm trackpy \citep{allan_2024_12708864}. Particle velocities are calculated using the inter-frame displacements and the timestamp in the metadata of the images. The number $N(t)$ of discharged particles over time is determined by counting the particles exiting the channel.

In addition to particle trajectories, some experiments require monitoring the total flow rate $Q$ within the channel during particle discharge. This measurement is particularly useful for estimating the system's hydraulic resistance $R_h=\Delta P/Q$ in pressure driven experiments. The use of a flow sensor is not possible here, as such a device would significantly increase the system's resistance to flow. Instead, two alternative measurement methods are employed in this study.

The first and most accurate method consists of measuring the fluid velocity in the channel using micro-PIV. After particle fabrication, the microfluidic channel is filled with tracer particles and transferred to a micro-PIV setup (Litron Lasers, Nano Series, $532$ nm). Fluid velocity is measured in a region far upstream of the orifice, where no particles are present. Due to the small depth of correlation in our micro-PIV setup, velocity measurements are confined to a $z$-plane of thickness $18.4$ \textmu m, smaller than the channel thickness \citep{olsen_out--focus_2000}. To calibrate the vertical position of the measurement plane within the channel, we first measure the fluid velocity at different $z$-positions during a steady flow and fit a parabolic velocity profile \citep{pimenta_viscous_2020}. The flow rate during particle discharged is then calculated by integrating the measured fluid velocity over the channel cross-section, assuming a parabolic profile in the vertical direction and a uniform velocity along the channel width.

While the micro-PIV method enables precise measurements of $Q$ over time, its use requires switching setups and calibrating the vertical position of the measurement plane before each experiment. Additionally, this method does not allow for simultaneous observation of the tapered section during particle discharge. For these reasons, we also use another method, where $Q$ is estimated by measuring the mass $m$ exiting the channel over time. In this method, both the fluid and particles leaving the channel are directed to a precision balance (Denver Instrument TP-214, $0.1$ mg accuracy, $1$ Hz) via a tubing of diameter $3$ mm, sufficiently large to prevent clog formation. The end of the tubing is immersed in a pool of solution to avoid dripping. The total flow rate $Q$ is directly deduced from mass measurements using $Q=\dot m / \rho$.

\subsection{Driving force}

\begin{figure}[tb]
    \begin{center}
        \includegraphics[width=0.76\linewidth]{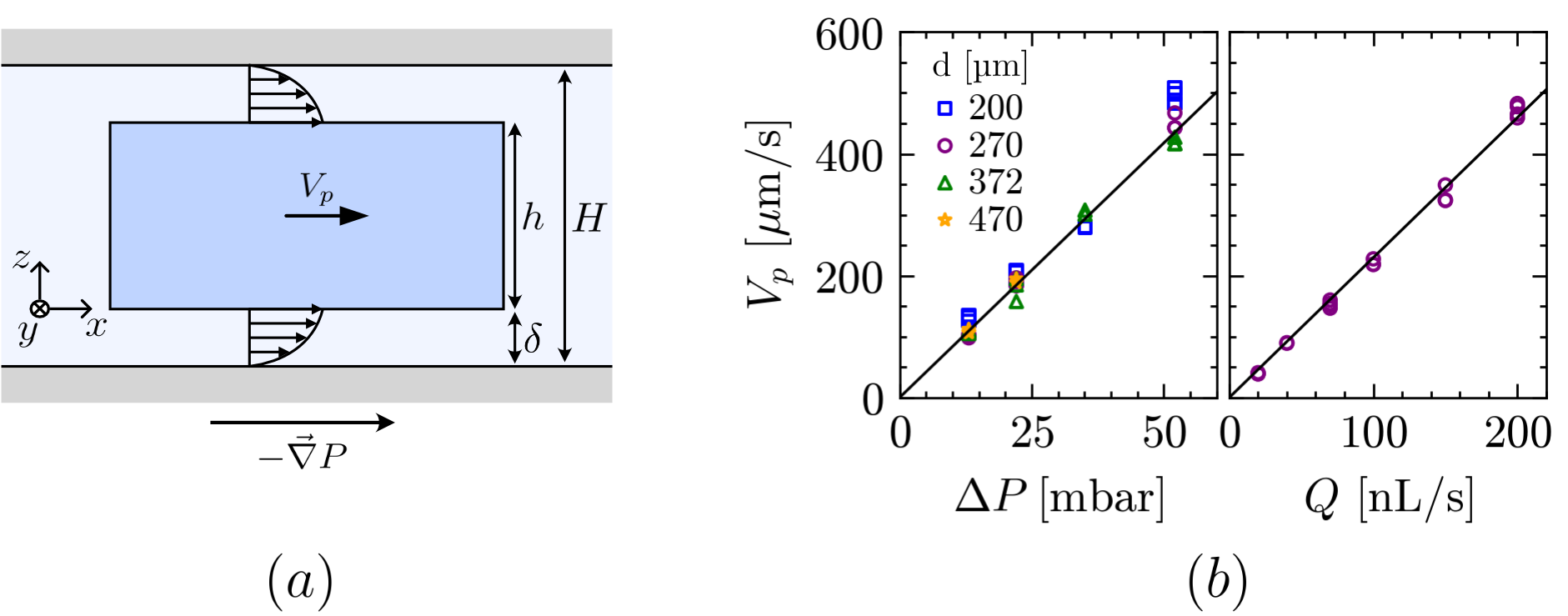}
        \caption{(a) Schematic side view of a particle with thickness $h$ inside a channel of height $H$, subjected to a pressure gradient $\vec \nabla P$. The particle moves with velocity $V_p$ along the $x$ direction, while the fluid flows through the gaps of thickness $\delta = (H - h)/2$, following a Poiseuille-Couette profile. (b) Measurements of $V_p$ in an empty channel for various particle diameters at different imposed pressure drops $\Delta P$ (left panel) and flow rates $Q$ (right panel). The black lines correspond to equation~\ref{eq:vp_q} without any adjusting parameter. In the pressure-imposed case, $Q$ is substituted by $\Delta P/R_{h,0}$, where $R_{h,0}$ is the empty channel resistance, measured prior to experimentation using micro-PIV.} 
        \label{fig:Vp_single}
    \end{center}
\end{figure}

A particle in the channel is subject to a pressure force due to the pressure gradient along the channel, a viscous drag force from the fluid flow around it, and a viscous friction force in the gaps above and below the particle \citep{berthet_single_2013}. For a disk-shaped particle of radius $r$ and height $h$ in a Hele-Shaw flow, as depicted in Figure~\ref{fig:Vp_single}(a), the pressure force and viscous friction are given by 
\begin{equation}
    \vec{F_p} = 2h\int_0^\pi P(\theta) r cos \theta \mathrm{d} \theta \vec{e_x} = \pi h r^2 \frac{\Delta P_p}{2r} \vec{e_x},
\end{equation}
\begin{equation}
    \vec{F_v} = 2\pi r^2 \left(\frac{\Delta P_p}{2r} \frac{\delta}{2} - \eta \frac{V_p}{\delta} \right) \vec{e_x},
\end{equation}
where $\delta$ is the gap thickness, $\Delta P_p$ is the decrease of pressure over the length of a particle and $V_p$ is the particle velocity. The viscous drag can be neglected as the fluid and particle velocities are very close to each other in our experimental geometry, where $2\delta \sim h$. The particle velocity can thus be determined by balancing the pressure force with the viscous dissipation in the gaps, yielding
\begin{equation}
    \label{eq:1}
    V_p = -\frac{\Delta P_p}{2r} \frac{\delta(h+\delta)}{2\eta}.
\end{equation}
This equation shows that particle motion in our system is driven by the pressure gradient over the particle length, $-\Delta P_p/2r$, mediated by a geometrical factor and the inverse of the viscosity.

The pressure drop $\Delta P_p$ can be expressed as $\Delta P_p = R_{h,p} Q$, where $Q$ is the total flow rate of the fluid and particles, and $R_{h,p}$ is the hydraulic resistance of the channel over the particle length. In the simple case of a single particle in the straight section of the channel, the particle's influence on the hydraulic resistance can be neglected, and $R_{h,p}$ can be approximated by the resistance of a slit channel with length $2r$, height $H$ and width $w$, expressed as $ R_{h,p} = 24\eta r / (H^3w)$. Substituting $\Delta P_p$ into eq~\ref{eq:1}, this gives
\begin{equation}
    V_p = \frac{6\delta(h+\delta)}{H^3 w} Q.
    \label{eq:vp_q}
\end{equation}
Equation~\ref{eq:vp_q} demonstrates that the velocity of a single particle in the channel is proportional to the flow rate $Q$, with a coefficient that depends only on the channel geometry and particle thickness. The validity of this proportionality is verified for varying particle diameters at different imposed flow rates and pressure drops, as shown in Figure~\ref{fig:Vp_single}(b). The black lines represent the predictions of equation~\ref{eq:vp_q} for $H=92$ \textmu m, $\delta=21$ \textmu m, $h=50$ \textmu m and $w=5$ mm. In the pressure-imposed case, $Q$ is replaced by $\Delta P/R_{h,0}$, where $R_{h,0}$ is the hydraulic resistance of the microfluidic system in the absence of particles, measured prior to experimentation using micro-PIV.

The driving mechanism in our system differs significantly from commonly studied granular hoppers, such as gravity- or conveyor belt-driven hoppers. While particles in these systems are driven by a uniform force, the driving force in our microfluidic hopper is dependent on the fluid velocity, which scales inversely with channel width, causing particles to accelerate as they pass through the constriction. Another difference with other systems lies in the temporal variations of the driving force during particle discharge. In most granular hoppers, the driving force remains constant as the channel empties. However, in our microfluidic system, the driving force depends on the overall hydraulic resistance of the channel, which might be influenced by the presence of particles. This complex interplay between fluid flow and particles could potentially lead to a reduction in hydraulic resistance as the hopper empties, resulting in a progressive increase of the driving force during the discharge process. Since the evolution of the outflow rate during the particle discharge is non-trivial, we first studied the discharge of particles from our microfluidic hopper over time.

\section{Particle discharge}

\subsection{Constant discharge rate and hydraulic resistance}

\begin{figure}[tb]
    \begin{center}
        \includegraphics[width= .9\linewidth]{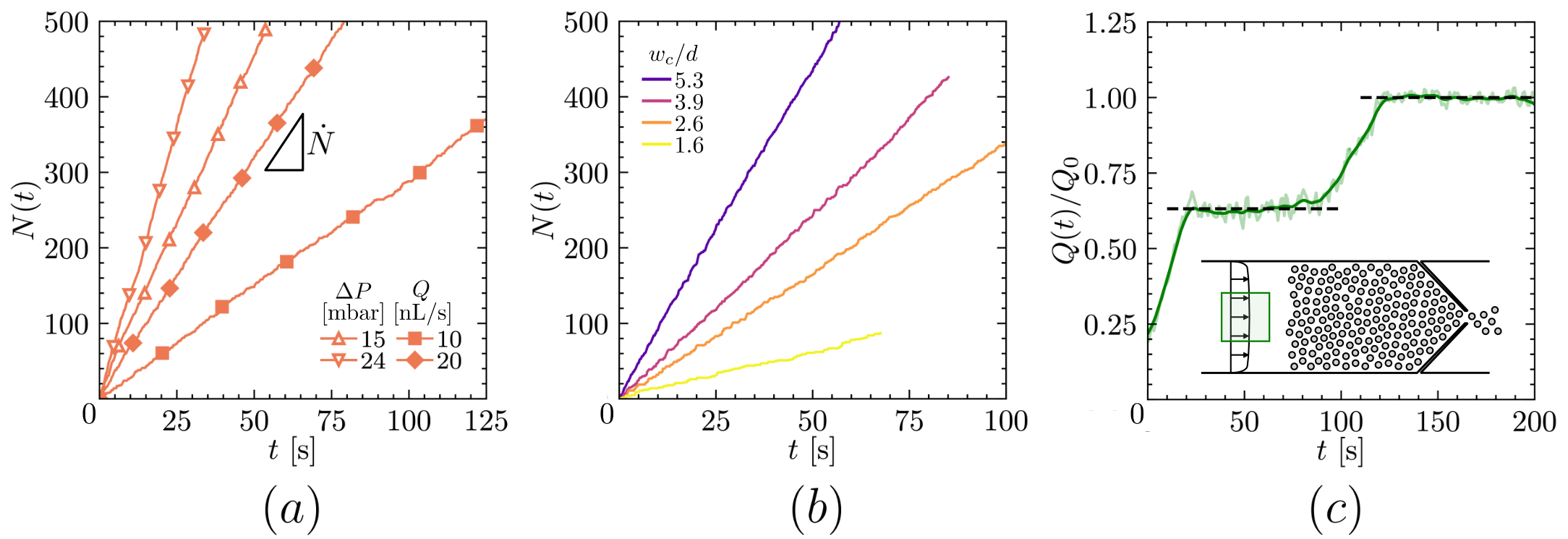}
        \caption{Discharge curves $N(t)$ for (a) different imposed flow rates $Q$ and pressure drops $\Delta P$ at fixed size ratio $w_c/d=3$, and (b) different orifice-to-particle size ratios $w_c/d$ under a fixed pressure drop $\Delta P=8$ mbar. All curves present a linear increase in the number of discharged particles $N$ over time. Note that the symbols are used to distinguish the datasets; data was collected at a frequency of $5$ Hz. Panel (c) presents micro-PIV measurements of the flow rate $Q(t)$ during a pressure-driven discharge with $\Delta P=15$ mbar. The flow rate $Q(t)$, deduced from fluid velocity measured in the region highlighted in green, is normalized by $Q_0$, its final value in an empty channel.} 
        \label{fig:cst_q}
    \end{center}
\end{figure}

We performed discharge experiments with varying particle diameters ($170$ \textmu m $\leq d \leq$ $400$ \textmu m), and two orifice widths ($w_c = 600$ \textmu m and $900$ \textmu m). In each experiment, we either imposed a flow rate, $Q$, or a pressure drop $\Delta P$, and measured the number $N(t)$ of particles discharged from the channel over time. Figure~\ref{fig:cst_q}(a) presents typical discharge curves for different imposed flow rates and pressure drops at fixed size ratio $w_c/d=3$. 

For both driving methods, we observe that $N$ scales linearly with time, indicating a constant discharge rate $\dot N$. This behavior is expected under imposed flow rate, as the force exerted by the fluid on particles remains constant throughout the experiment. However, the constant discharge rate under imposed pressure is more surprising, as one would expect the hydraulic resistance $R_h$ to decrease as the channel empties, leading to a progressive increase of the flow rate. This, in turn, should result in an increasing particle discharge rate, as reported for submerged granular flows \citep{koivisto_sands_2017, wilson_granular_2014}. Note that the constant discharge rate under imposed pressure conditions is observed regardless of the orifice-to-particle size ratio within the range of diameters considered, as seen in Figure~\ref{fig:cst_q}(b). Even in the case of frequent clog formation ($w_c/d=1.6$), a constant discharge rate is observed between clogging events.

The constant discharge rate under imposed pressure conditions suggests that the hydraulic resistance $R_h$ of our experimental system remains unchanged throughout each experiment, despite the reduction in the number of particles within the channel. To verify this, we measured $Q(t)$ during a pressure-driven discharge experiment ($\Delta P = 15$ mbar) using micro-PIV. In this configuration, particles are not observable, and their number within the channel remains unknown. 

Figure~\ref{fig:cst_q}(c) displays the temporal evolution of $Q(t)/Q_0$ throughout an experiment, with $Q_0$ representing the final flow rate measured after full discharge. The valve is opened at $t=0$, and after a brief period ($\sim 20$ s) needed to initiate particle motion, the flow rate stabilizes at approximately 60\% of $Q_0$ for about a minute, consistent with the typical duration of a discharge under these conditions (see Figure~\ref{fig:cst_q}(a)). Eventually, the channel completely empties and the flow rate reaches $Q_0$. Since $R_h = \Delta P/Q$, this result confirms that the hydraulic resistance remains constant during most of the discharge process. The $\sim40$\% decrease in $Q$ during particle discharge further indicates that $R_h$ is not solely controlled by a predominantly high channel resistance, but is also governed by the presence of particles within the channel.

\subsection{Hydraulic resistance controlled by particles at the orifice}

\begin{figure}[tb]
\centering
    \includegraphics[width = \linewidth]{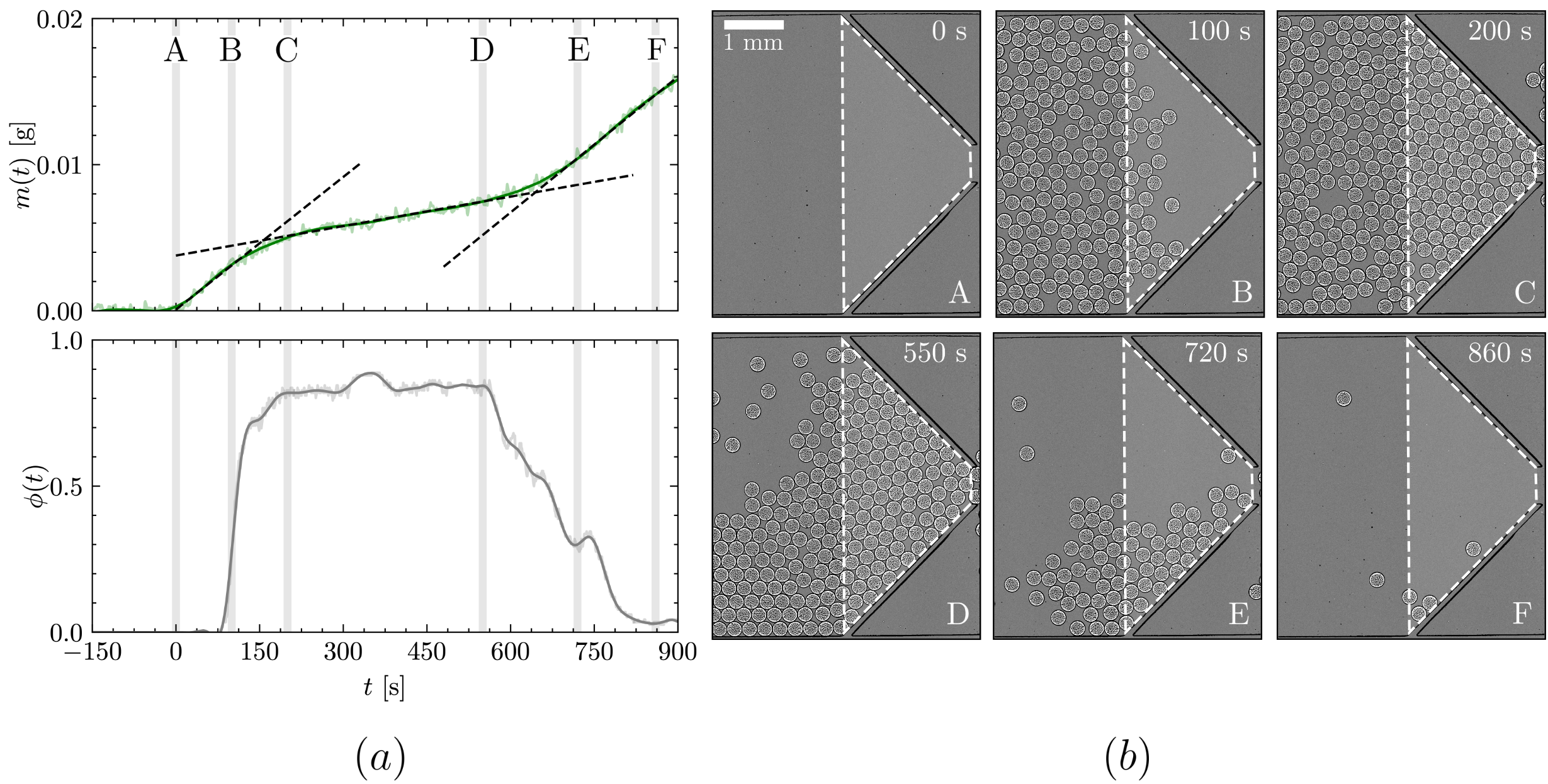}
    \caption{(a) Temporal evolution of the discharged mass $m(t)$ and the constriction solid fraction $\phi(t)$ during the flow of a dense packing through an empty constriction. The gray vertical lines correspond to the snapshots of the channel shown in (b), labeled from A to F. Measurements of $\phi(t)$ are taken in the area outlined in white in the snapshots.}
    \label{fig:rh_hopper_sec}
\end{figure}

While the previous experiments clearly demonstrate a constant outflow rate and hydraulic resistance during discharge, attributing this constant flow to the presence of particles is difficult, as we cannot observe particles at the orifice during micro-PIV measurements. To overcome this limitation, we conducted an additional experiment where we observed the flow of a dense particle packing through an empty constriction under imposed pressure conditions, while monitoring the mass of particles and fluid leaving the channel to measure $Q$. 

In this experiment, around $1000$ particles were first packed at the orifice, and then gently transported backward by a slight reverse flow, as to not disturb the packing. Once the packing was sufficiently distanced from the orifice ($\sim5$ mm), the flow was stopped by imposing zero pressure ($t<0$). At $t=0$, a pressure drop of $\Delta P = 8$ mbar was applied, driving the particles back towards the constriction. As the packing approached and passed through the tapered section, both the mass $m(t)$ exiting the channel and the solid fraction $\phi(t)$ in the constriction were measured. Mass and solid fraction measurements are presented in Figure~\ref{fig:rh_hopper_sec}(a), while the corresponding snapshots of the channel, labeled from A to F, are shown in Figure~\ref{fig:rh_hopper_sec}(b). The region used to measure $\phi(t)$ in the constriction is highlighted in white on each snapshot.  

Three distinct phases can be identified, each with a specific, constant flow rate, as indicated by the black dashed lines representing linear fits of $m(t)$ in Figure~\ref{fig:rh_hopper_sec}(a). During the first phase, the empty constriction (snapshot A) fills with particles as the dense packing enters the constriction (snapshot B), with a fitted flow rate of $Q_1\approx 27.0$ nL/s. Once the tapered section becomes completely filled with particles, the solid fraction $\phi$ reaches a maximum value of $\phi\approx0.83$ (snapshots C and D). During this second phase, $\phi$ remains constant, except around $t=350$ s, when a transient clog forms, slightly compacting the particles further. The second phase is characterized by a significantly reduced flow rate, with a fitted value of $Q_2\approx 6.0$ nL/s $\approx0.2Q_1$. Finally, the number of particles within the channel becomes insufficient to completely fill the constriction (snapshot E), and $\phi$ decreases until nearly no particles remain in the channel (snapshot F). During this final phase, the flow rate rises back to $Q_3\approx 27.6$ nL/s $\approx1.02Q_1$.

These results indicate that the flow rate $Q$ and hydraulic resistance $R_h$ during particle discharge are primarily controlled by the presence of densely packed particles within the tapered section, regardless of the total number of particles in the channel. This suggests that most of the energy dissipation arises from the dense packing being squeezed through the constriction, with the dissipation occurring either through particle rearrangements or by viscous dissipation due to large velocity fluctuations. While the exact origin of this dissipation in the tapered section remains to be understood, the constant hydraulic resistance and the subsequent constant discharge rate allow for a comparison with other commonly studied hopper flows. In the next two sections, we will show that our experimental system recovers similar scaling laws for the outflow and clogging of hard monodisperse disks through a constriction as observed in other dry and suspension systems.

\section{Applications}

\subsection{Beverloo-like discharge}

As discussed previously, the flow of particles through a bottleneck under gravity is usually described by the Beverloo law. In the 2D case, this law is commonly expressed as
\begin{equation}
    q_p = C \phi \sqrt{g} (w_c - k d)^{1.5},
    \label{eq:beverloo}
\end{equation}
where $q_p$ is the particle volumetric discharge rate, $\phi$ the particle solid fraction and $g$ the acceleration due to gravity, while $k$ and $C$ are two fitting parameters. The Beverloo law is based on the assumption that the discharge of particles through a constriction is limited by (i) an effective outlet size, which accounts for a no-flow zone along the orifice margins, and (ii) the exit velocity of particles within the orifice. The effective outlet size is expressed as $w_c - kd$, where $k$ is a fitting parameter which depends on particle shape and hopper slope, with typical values between 1 and 3 \citep{beverloo_flow_1961, mankoc_flow_2007}. The exit velocity at the outlet depends on the driving force of the system. In the gravity-driven case, this velocity corresponds to that of a particle falling from an arch of height comparable to the effective orifice size, resulting in the term $\sqrt{g} \, (w_c-kd)^{0.5}$ in equation~\ref{eq:beverloo} \citep{beverloo_flow_1961, dorbolo_influence_2013}. In contrast, particles transported on a conveyor belt exit the outlet region at the belt speed ($q_p \propto V_{belt}$) \citep{aguirre_pressure_2010, de-song_critical_2003}, while the exit velocity of bubbles rising in an inverted silo is set by buoyancy and drag forces ($q_p \propto g^{1.5}$) \citep{bertho_dense_2006}.

\begin{figure}[tb]
\centering
    \includegraphics[width = .95\linewidth]{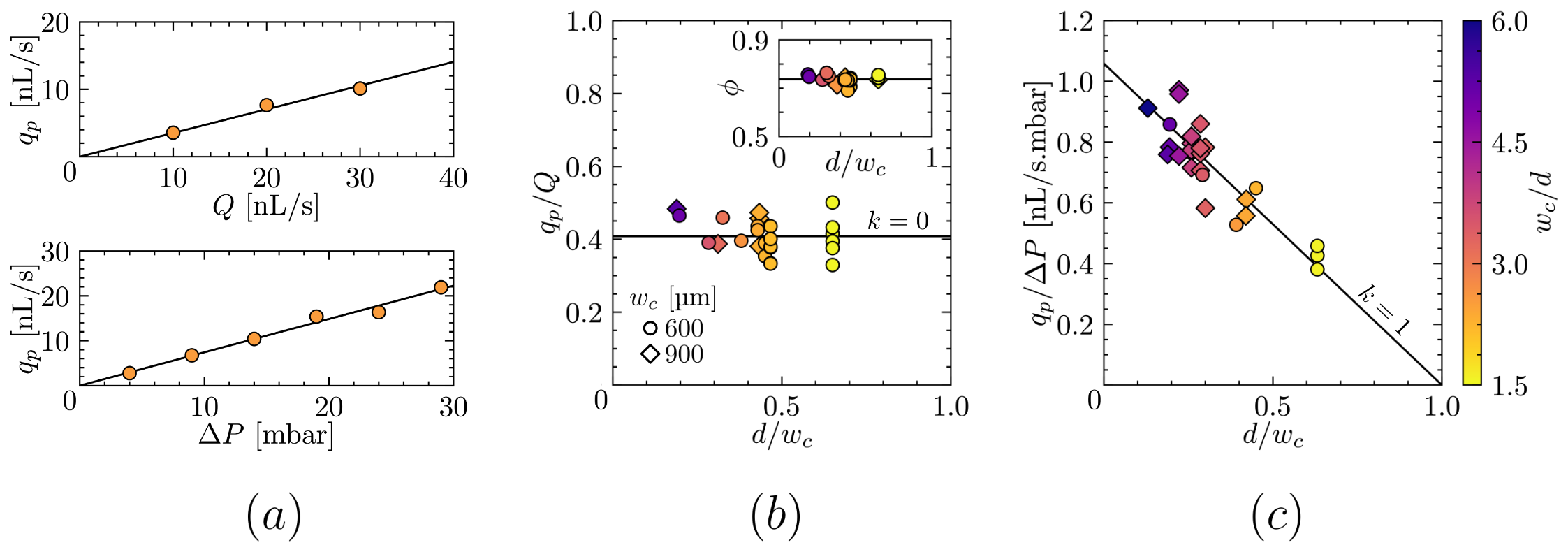}
    \caption{(a) Particle discharge rate, $q_p$, measured for a fixed size ratio $w_c/d=2.5$ under various imposed flow rates $Q$ (top panel) and pressure drops $\Delta P$ (bottom panel). Solid lines indicate linear fits. (b) and (c) Particle discharge rate $q_p$ divided by the total flow rate $Q$ and pressure drop $\Delta P$, respectively, as a function of $d/w_c$. The solid line represents the fit of equation~\ref{eq:scaling_predicted}. Variations of the solid fraction near the orifice $\phi$ with the size ratio $d/w_c$ are shown in the inset of (b).}
    \label{fig:beverloo}
\end{figure}

To investigate the relationship governing particle flow rate in our system, we performed discharge experiments with varying driving intensities, orifice widths, and particle diameters. In each experiment, we determined the discharge rate $\dot N$ by fitting the slope of the particle outflow over time, $N(t)$, and deduced the volumetric particle discharge rate $q_p = \dot N \pi h d^2/4$.
We first examined the influence of the driving force on particle discharge by applying different flow rates $Q$ or pressure drops $\Delta P$, while keeping the outlet and particle sizes constant ($d=237$ \textmu m, $w_c=600$ \textmu m). Figure~\ref{fig:beverloo}(a) shows a linear scaling of $q_p$ with the driving intensity for both driving mechanisms, consistent with the expression for the velocity of isolated particles given in equation~\ref{eq:vp_q}. This linear relationship indicates that particles rapidly reach their terminal velocity in the exit region, as the flow is dominated by viscous forces \citep{guariguata_jamming_2012}.

We also investigated the influence of orifice and particle dimensions on particle outflow under both imposed flow rate and pressure conditions. To isolate the effect of geometry from that of the driving intensity, we measured $q_p/Q$ and $q_p/\Delta P$ for varying particle diameters $d$ and constriction widths $w_c$. By mass conservation of the driving fluid, the exit velocity of particles in our system is expected to scale inversely with $w_c$. Considering an effective orifice width of $w_c-kd$, the Beverloo model predicts the following scaling:
\begin{equation}
    \label{eq:scaling_predicted}
    q_p\propto 1 - kd/w_c.
\end{equation} 
Figures~\ref{fig:beverloo}(b) and \ref{fig:beverloo}(c) show the evolution of $q_p/Q$ and $q_p/\Delta P$ as functions of $d/w_c$, respectively. In both figures, the solid black line represents the fit of the parameter $k$ obtained from equation~\ref{eq:scaling_predicted}. Surprisingly, the two curves show different results. Under imposed flow rate, we find $k \approx 0$, indicating that $q_p$ is mostly independent of both particle and orifice sizes. In contrast, the pressure-driven discharge aligns well with the Beverloo scaling, with a fitted value of $k$ close to $1$.

The observation that $k\approx0$ under imposed flow rate suggests that no correction for particle dilution at the orifice is needed in our microfluidic hopper \citep{mankoc_flow_2007}. We propose that the volumetric particle discharge rate in our system is given by 
\begin{equation}
    \label{eq:qp_Q}
    q_p = \phi h w_c V_t,
\end{equation}
where $\phi$ is the particle solid fraction near the orifice, $h$ the particle thickness, and $V_t$ the exit velocity, scaling as $Q/w_c$. Since $\phi$ remains nearly constant in our experiments ($\phi\approx0.74$, see inset of Figure~\ref{fig:beverloo}(b)), equation~\ref{eq:qp_Q} leads to a constant ratio $q_p/Q$ independent of the size ratio $d/w_c$, consistent with experimental observations. In addition, the experimentally measured ratio $q_p/Q \approx  0.41$ is well captured by substituting the exit velocity $V_t$ with the expression given by equation~\ref{eq:vp_q}, yielding $q_p = \phi 6h\delta(h+\delta )H^{-3}Q \approx 0.43 Q$.

Unlike the flow rate-driven case, the discharge of particles under a pressure gradient $\Delta P$ follows a Beverloo-like scaling and can be described by $q_p \propto \Delta P (1 - d/w_c)$. The difference between flow rate- and pressure-driven scenarios may arise from variations of the hydraulic resistance $R_h$ with $w_c$ and $d$ during pressure-driven discharge. Indeed, energy dissipation in the tapered section of the channel could vary depending on both the orifice and particle sizes. The exact coupling between particle discharge and hydraulic resistance, as well as the origin of the Beverloo-like scaling, remain to be understood.

\subsection{Clogging}

\begin{figure}[tb]
\centering
    \includegraphics[width = .87\linewidth]{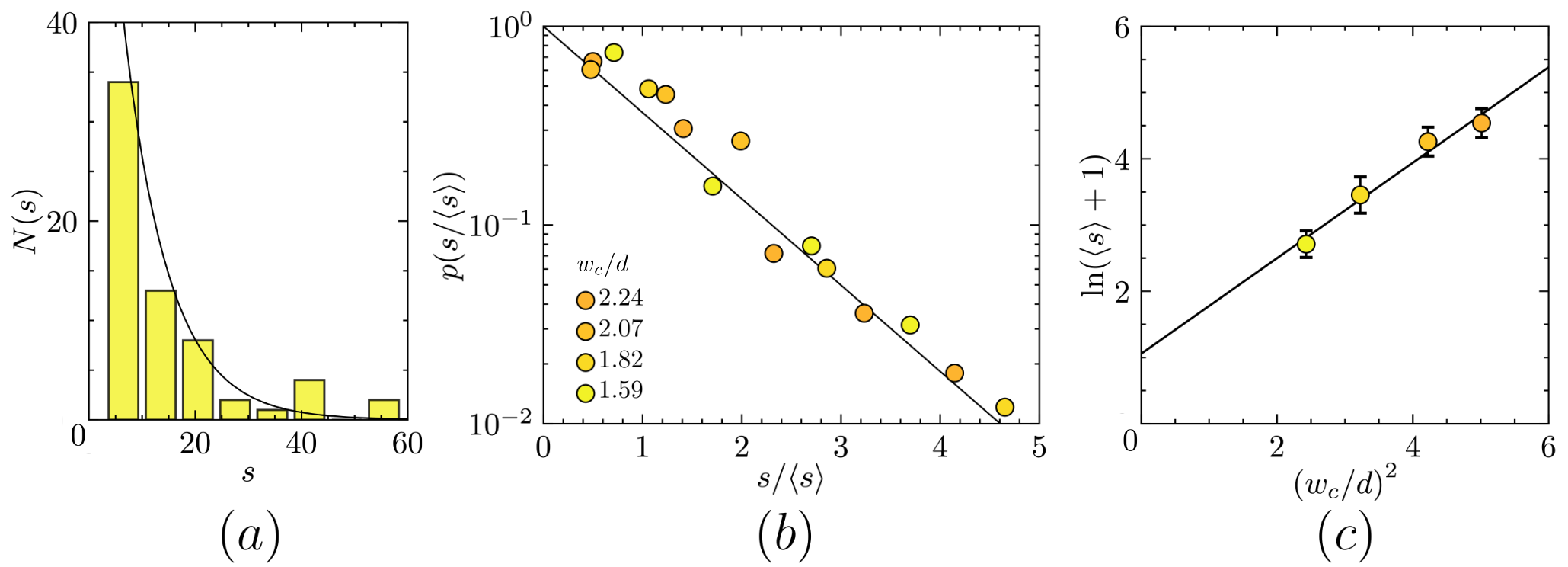}
\caption{(a) Histogram showing the number of events $N$ in which $s$ particles escape the channel before clog formation, for a fixed size ratio $w_c/d=1.59$. The solid line represents an exponential distribution. (b) Probability distribution function of the normalized avalanche size $s/\langle s \rangle$ for all size ratios considered, fitted with an exponential law. (c) Plot of $ln \left( \langle s \rangle + 1\right)$ as a function of $\left( w_c/d\right)^2$. The solid line is a linear fit, and error bars are obtained using a bootstrapping method. All experiments were performed under a fixed pressure drop $\Delta P = 18$ mbar.}
\label{fig:clogging}
\end{figure}

We now examine the clogging of particles in our experimental setup. Clogging is usually characterized by the number of particles, $s$, exiting the channel before clog formation, commonly referred to as the avalanche size. Both experimental \citep{marin_clogging_2024} and theoretical \citep{helbing_analytical_2006} studies have widely demonstrated that the avalanche size distribution follows an exponential decay. This behavior arises from the stochastic nature of clogging, where each particle exiting the channel has an equal probability $p_{clog}$ of forming a clog, resulting in a geometric distribution of the avalanche size probability: 
\begin{equation}
    p(s) = p_{clog} (1-p_{clog})^s.
\end{equation} 
The average avalanche size, $\langle s \rangle$, which is related to the clogging probability by $\langle s \rangle = p_{clog}/(1-p_{clog})$, strongly depends on the ratio of the orifice width $w_c$ to the particle diameter $d$. In 2D gravity-driven systems, $\langle s \rangle$ can be described by an exponential-square growth with $w_c/d$, expressed as
\begin{equation}
    \langle s \rangle = A e^{B \left( w_c/d\right) ^{2}} - 1,
    \label{eq:clog}
\end{equation}
where $A$ and $B$ are two constants \citep{janda_jamming_2008, to_jamming_2005}.

In this section, we verify that our microfluidic system recovers the same scaling laws for clogging as observed in other systems. To this end, we conducted discharge experiments using disk-shaped particles with diameters ranging from $268$ to $378$ \textmu m. Particles were discharged through a constriction of width $w_c=600$ \textmu m, resulting in orifice-to-particle size ratios $w_c/d$ between 1.59 and 2.24. The discharge experiments were pressure-driven ($\Delta P = 8$ mbar) rather than flow rate-driven, as pressure controllers have a faster response time than syringe pumps. We considered that the channel was clogged when an arch of particles formed at the constriction and persisted for more than 3 seconds. Avalanches with fewer than four particles were excluded, as they may involve pre-existing clogging arches. 

For each particle size considered, the avalanche size distribution $N(s)$ was determined by measuring $s$ over a large number of clogs. Each distribution can be described by an exponential fit, as indicated by the black line in Figure~\ref{fig:clogging}(a) for $w_c/d=1.59$. We verified that all distributions collapsed on the same exponential law when normalized by the average avalanche size (Figure~\ref{fig:clogging}(b)), consistent with prior studies. Additionally, we show in Figure~\ref{fig:clogging}(c) that the evolution of $\langle s \rangle$ with $w_c/d$ in our experimental setup is consistent with the scaling of equation~\ref{eq:clog}, with $\ln{\left(\langle s \rangle + 1 \right)}$ increasing linearly with $\left( w_c / d\right)^2$ across the range of orifice-to-particle size ratios considered. Note that $\langle s \rangle$ might have been slightly underestimated for the smallest particle sizes (largest $w_c/d$), as large avalanches were limited in our experiments by the number of particles in the channel. This limitation could be addressed by using longer channels filled with more particles, or by instead measuring the fraction of experiments in which a clog forms.

\subsection{Perspectives}

In this paper, we focused exclusively on the flow of hard, monodisperse disk-shaped particles to characterize our experimental system and compare it to other well-established systems. However, the versatility of our experimental setup makes it an excellent model system for exploring more complex scenarios. In the following paragraphs, we outline perspectives for future works on hopper flow, and more broadly, in any 2D flow configuration.

\begin{figure}[tb]
\centering
    \includegraphics[width = .95\linewidth]{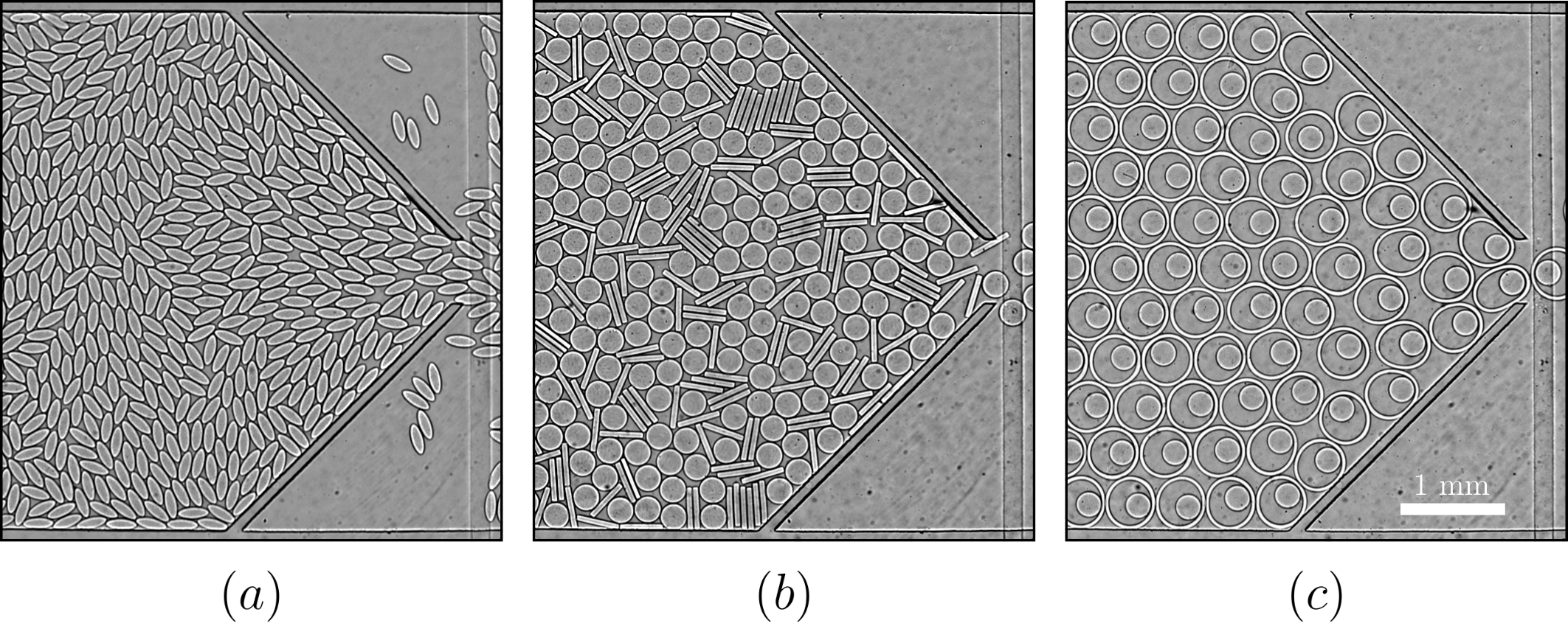}
    \caption{Examples of discharge experiments with (a) elongated grains, (b) a mixture of disks and rigid fibers, and (c) disks embedded within flexible rings (see Supplementary Movie 5).}
    \label{fig:demo}
\end{figure}

The shape of particles can be easily controlled in our experimental setup by changing the design of the UV mask used during particle fabrication. This method provides a fast and effective way of varying particle shape, opening new perspectives for studying the flow dynamics of non-circular grains. The fine and rapid control over particle shape enables systematic studies of shape properties through gradual changes, such as by precisely adjusting the aspect ratio of elongated particles, as seen in Figure~\ref{fig:demo}(a).
Since there is no flow during the fabrication process, particles with different shapes can also be combined by first printing a batch of particles, then changing the UV mask and printing another batch in the remaining spaces. This method can be applied to fabricate circular particles with finely controlled polydispersity, or even mixtures of particles with different shapes, as illustrated in Figure~\ref{fig:demo}(b) for a disk-fiber mixture.

In addition to particle shape, our experimental setup provides a relatively precise way to modulate particle deformability, either by changing the composition of the crosslinking solution \citep{duprat_microfluidic_2014}, or by fabricating thin, slender structures that can deform under flow \citep{cappello_transport_2019}. More complex situations combining particles of different deformability can also be achieved, as illustrated in Figure~\ref{fig:demo}(c), where hard disks are embedded within flexible rings (see Supplementary Movie 5).
Although all the examples provided involve dense packings, dilute configurations can also be achieved with our experimental system, as the in-situ fabrication method provides precise control over the spatial distribution of particles before flow.

\section{Conclusion}

In this paper, we presented a two-dimensional microfluidic setup designed to investigate the flow and clogging of particles in confined geometries. This innovative setup relies on an in-situ particle fabrication method, which enables easy and precise control over the particle shape, deformability and volume fraction. Experiments were conducted in a microfluidic hopper channel, where a Quake valve positioned immediately after the constriction was used to concentrate particles up to the maximum packing fraction. Our study focused on the flow of a dense packing of hard, monodisperse, disk-shaped particles through a bottleneck under both imposed flow rate and pressure conditions. 

We have shown that, despite fundamental differences between our microfluidic system and typical dry granular systems, particles discharge at a constant rate under both imposed flow rate and pressure conditions in our experimental setup. This constant particle outflow rate arises from the fact that the hydraulic resistance is constant during the entire discharge, and dominated by the presence of densely packed particles within the tapered section of the hopper. We also demonstrated by measuring the particle discharge rate for varying driving intensities, particle diameters and constriction widths that the outflow of particles in our system depends on the flow driving mechanism. While the particle discharge rate is mostly independent of orifice and particle sizes under imposed flow rate conditions, the pressure-driven discharge of particles in our system is well described by a Beverloo-like scaling. Finally, we showed that the statistics of clog formation in our experimental system follow the same stochastic laws as reported in other systems.

The good agreement between our experimental results and other commonly studied systems proves its utility for studying the discharge and clogging of particulate suspensions in various two-dimensional configurations. The versatility of our microfluidic system allows for future investigation of the role of particle shape, size distribution and deformability on the outflow and clogging of suspended particles through a bottleneck.

\section{Acknowledgements}

LK and AL acknowledge funding from the European Union’s Horizon 2020 Research and Innovation Program CALIPER under the Marie Skłodowska-Curie grant agreement number 812638. JT acknowledges funding from ENS de Lyon through a CDSN grant. We thank Institut Pierre-Gilles de Gennes (Investissements d’avenir ANR-10-EQPX-34)

\bibliographystyle{apalike}
\bibliography{bibli, Bib2}
\setlength{\itemsep}{0pt}  
\end{document}